\documentclass[twocolumn,a4paper]{article}

\input{epsf.tex}

\newcommand{\be}{\begin{equation}}
\newcommand{\ee}{\end{equation}}
\newcommand{\ba}{\begin{eqnarray}}
\newcommand{\ea}{\end{eqnarray}}
\newcommand{\ban}{\begin{eqnarray*}}
\newcommand{\ean}{\end{eqnarray*}}

\newcommand{\ket}[1]{\mbox{$ | #1 \rangle $}}

\newcommand{\PSpetitGr}[2]{\def\epsfsize##1##2{#2\textwidth} \vspace{10pt}
\centerline{\epsfbox{#1}} \vspace{10pt}}

\begin{document}

\title{\large\bf Does entanglement depend on the timing of the impacts at the 
beam-splitters?}

{\normalsize{\author{{\bf Antoine Suarez}\thanks{corresponding author; suarez@leman.ch}\\
Center for Quantum Philosophy\\
The Institute for Interdisciplinary Studies\\
P.O. Box 304, CH-8044 Zurich, Switzerland
\and
{\bf Valerio Scarani}\thanks{valerio.scarani@ipe.dp.epfl.ch}\\
Institut de Physique Exp\'erimentale\\
Ecole Polytechnique F\'ed\'erale de Lausanne\\
PHB-Ecublens, CH-1015 Lausanne, Switzerland
}}}
\maketitle

{\small A new nonlocality experiment with moving beam-splitters is proposed. 
The experiment is analysed according to conventional quantum 
mechanics, and to an alternative nonlocal description in which 
superposition depends not only on indistinguishability but also on 
the timing of the impacts at the beam-splitters.

}

\section{Introduction}
{\em Entanglement} is Schr\"{o}dinger's name for superposition in a 
multi-particle system. In particular he called {\em entangled states} 
two-particle states that cannot be factored into products of two 
single-particle states in any representation. Multi-particle 
superposition is considered to be the characteristic trait of quantum 
mechanics \cite{ghz93}. If one cannot distinguish (even in principle) 
between different paths from source to detector, the amplitudes for 
these alternative paths add coherently, and multi-particle 
correlations appear. If it is possible in principle to distinguish, 
correlations vanish.
Work by John Bell \cite{bel64}, by Daniel M. Greenberger, Michel A. Horne, 
and Anton Zeilinger \cite{ghz89}, and by Lucien Hardy \cite{har93} pointed out that 
local-realistic theories cannot account for the two- and multiparticle 
correlations implied by the superposition principle. Local-realism 
is the name for Einstein's assumption that nothing in physical 
reality happens fasten than light. The quantum mechanical violation of 
local-realism is now mostly called nonlocality. In spite of the 
loopholes in the experiments \cite{div}, it is today largely accepted 
that superluminal nonlocality is a feature of nature: most 
physicists will not be surprised, if a future "loophole free" 
Bell experiment \cite{kwi94} definitely demonstrates the violation of 
the locality criteria (Bell's inequalities or others). Nevertheless 
nonlocality cannot be used by human observers for practical purposes 
(impossibility of "superluminal signaling"). Hence, if one accepts 
that relativity experiments like Michelson-Morley only imply a 
practical impossibility for man to communicate faster than light, 
no contradiction between these experiments and quantum mechanics 
arises.

However, the heart of relativity is the conclusion that there is no 
absolute spacetime, no "aether". Simultaneity depends on the 
observer's state of movement, the order of succession of two spacelike separated 
events may change if one changes the inertial frame. Relativity of 
space-time seems to be at odds with the quantum mechanical assumption 
that superposition depends exclusively on indistinguishability 
\cite{ghz93,kwi94,fra89}, 
and not on the times at which the values are measured, in any 
inertial frame whatsoever. Consider for instance the orthodox quantum 
mechanical description of the perfect EPR correlations in two-particle 
experiments with entangled polarized photons: according to the 
superposition principle, the spin operator related to a measuring 
apparatus with two parallel oriented polarizing beam-splitters has 
two eigenvectors $\ket{+1,+1}$ and $\ket{-1,-1}$, representing two orthogonal quantum 
eigenstates; the measurement causes the entangled state to jump into either the 
state $\ket{+1,+1}$ or the state $\ket{-1,-1}$ instantaneously, where the first 
state means that both photons are detected in the detectors monitoring the 
transmitted output ports, and the second one that both photons are detected in the 
 detectors monitoring the reflected output ports. Consequently, 
 the measurement produces events which are simultaneously 
 strictly correlated in spacelike separated regions. But 
 in which inertial frame are these correlated events 
 simultaneous? Quantum mechanics does not answer 
 this question. Moreover, because each measurement of polarization 
 may lie outside the other's light cone (e.g. the two measurements 
 may be spacelike separated events), the measurement which is 
 considered as the cause of the "jump" in a certain inertial frame, 
 is no longer the cause in another inertial frame. For one observer 
 the value measured at side 1 depends on which value has been 
measured at side 2, and for another observer the value measured at side 2 
 depends on which value has been measured at side 1. Different 
 observers are led to contradictory descriptions of the same 
 reality. That is the resason why there still seems to be no 
 consistent relativistic interpretation of the "quantum jump" 
 (also referred to as "reduction of the wave packet" or "wavefunction 
 collapse") associated with the measurement process, or why the 
notion of collapse appears to have no meaning in a relativistic context 
\cite{per95}. 
 Such causality paradoxes have also motivated models that give up 
 the relativity of spacetime and assume an absolute order of 
 succesion or "quantum aether" \cite{boh}.

Besides this conflict between relativity and quantum mechanics, we 
would like to highlight another problem more intrinsic to quantum mechanics 
itself. No matter if one accepts the relativity of space-time or 
quantum aether, the superposition principle seems to contradict somewhat 
another main postulate of orthodox quantum mechanics, namely "no values prior 
to measurement". According to this postulate, only after a specific 
measurement has been made can we attribute a definite physical 
property to a quantum system: there are no pre-existing values prior to the 
measurement, or, what is the same, the measurement creates the 
values. Therefore, the appearance of perfect nonlocal correlations 
necessarily express a link existing between {\em real} measured values, and one is led 
to assume that the measurement at one of the regions produces either a 
value $+1$ or $-1$ after taking into account the value that has {\em 
actually} been measured in the other region. This means that there is an order of 
succession, and thus the perfect correlations should disappear if 
the measurement in region 1 occurs, in the inertial frame of the 
analysing device in this region, {\em before} the measurement in region 2, 
and the measurement in region 2 occurs, in the inertial frame of the 
analysing device in this region, {\em before} the measurement in region 1. 
But even if one assumes a universal order of succesion and rejects 
the possibility two perform two "before" measurements, one has to 
admit the possibility of simultaneous measurements. For such a case it is 
absurd to assume {\em together} that photon 1 impacting at beam-splitter 1 
chooses the output port taking account of the choice photon 2 has 
{\em really} made at beam-splitter 2, and photon 2 chooses taking account 
of the choice photon 1 has {\em really} made. Therefore, in the case of 
simultaneous measurements in absolut space-time, the perfect correlations 
should also disappear. The superposition principle looks, therefore, 
to be at odds not only with relativity, but also with the postulate 
of "no values prior to the measurement". 

The purpose of this letter is to stress that quantum mechanics (QM in 
what follows) is a particular view of the relationship existing between 
superposition and indistinguishability, but that other views are possible. 
QM considers indistinguishability to be a {\em sufficient} condition 
for superposition. However, at the present time exclusively 
non-relativistic experiments without moving devices have been done. 
Stricktly speaking, such experiments support only the view that 
indistinguishability is a {\em necessary} condition for superposition. On 
this line of thinking we present in the following the basic features 
of an alternative nonlocal description which assumes that 
superposition does not depend exclusively on indistinguishability but 
also on the timing of the impacts at the beam-splitters. Furthermore 
we propose a relativistic experiment that may allow us to decide 
between this alternative view and QM.

\section{Definitions and Principles of the Alternative Description 
(AD)}

Consider the experiment with polarized photons sketched in Fig.1. 
Two classes of photon pairs, $(H_{1},H_{2})$ and $(V_{1},V_{2})$, are prepared 
through down-conversion in the "Bell state":
\be
\ket{\phi}=\frac{1}{\sqrt{2}}(\ket{H_{1},H_{2}}-\ket{V_{1},V_{2}})
\ee
where $H$ and $V$ indicate horizontal and vertical polarization, 
respectively. The polarizing beam-splitters $BS_{1}$ and $BS_{2}$ are vertical 
oriented, and preceeded by half wave plates, which rotate the 
polarization of the photons by angles $\alpha$, $\beta$. Suppose each beam-splitter 
can move fast, and change from one inertial frame to another.

\begin{figure}[ht]
\PSpetitGr{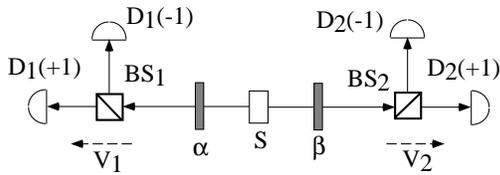}{0.4}
{\small\it{\caption{Experiment with polarized photons involving moving BS.}}}
\label{fig:setup}
\end{figure}

The proposed AD is based on the following 
definitions and principles:

If it is in principle impossible to know to which input sub-ensemble 
of $BS_{i}$, $i\in\{1,2\}$, a particle belongs by detecting it after 
leaving $BS_{i}$, then the impact 
at $BS_{i}$ is referred to as originating indistinguishability or uncertainty, 
and labeled $u_{i}$. If it is in principle possible to know to which input 
sub-ensemble of $BS_{i}$ a particle belongs by detecting it after leaving $BS_{i}$, 
then the impact at $BS_{i}$ is referred to as making possible 
distinguishability, and labeled $d_{i}$.

At the time $T_{i}$ at which a particle $i$ arrives at $BS_{i}$, we 
consider whether in the inertial frame of this beam-splitter, 
particle $j$ ($j\in\{1,2\}, j\neq i$) has already made a $u_{j}$ 
impact or not, and introduce the following definitions:

{\em Definition 1}: the impact of particle $i$ in $BS_{i}$ is a {\em 
before} event $b_{i}$ if:
\begin{enumerate}
\item{it is a $u_{i}$, and}
\item{either the impact of particle $j$ at $BS_{j}$ is a $d_{j}$ one; or 
$(T_{i}<T_{j})_{i}$, the subscript $i$ after the parenthesis meaning 
that all times referred to are measured in the rest frame of $BS_{i}$.}
\end{enumerate}

{\em Definition 2}: the impact of particle $i$ in $BS_{i}$ is a {\em 
non-before} event $a_{i}$ if:
\begin{enumerate}
\item{it is a $u_{i}$, and}
\item{the impact of particle $j$ at $BS_{j}$ is a $u_{j}$ one, and}
\item{$(T_{i}\geq T_{j})_{i}$.}
\end{enumerate}

We can now state the two principles of AD:

{\em Principle I}: if the impact of a photon $i$ at $BS_{i}$ is a 
$b_{i}$ impact, then photon $i$ produces values taking into account 
only local information, i.e., it is not influenced by the parameters 
photon $j$ meets at the other arm of the setup.

{\em Principle II}: if the impact of a photon $i$ at $BS_{i}$ is a 
$a_{i}$ impact, then photon $i$ takes account of photon $j$ in such a 
way that the values photon $i$ actually produces, and the values 
photon $j$ produces in a $b_{j}$ impact are correlated according to 
the superposition principle.

Assuming {\em Principle I} and {\em Principle II} we implicitly {\em 
discard} the hypothesis 
that the values produced by a particle, say photon 1, depend on the 
state of movement of the detectors $D_{1}$. Effective, the experimental data 
suggest that the outcome distribution does not depend on the distances 
at which the detectors are placed with respect to the beam splitters 
\cite{com10}. 
This is also in accord with the quantum mechanical formalism. It is 
reasonable, therefore, to assume that the value produced by photon 1, 
if it is detected after leaving splitter $BS_{1}$ and there is no other 
splitter between $BS_{1}$ and the detector, is determined at the time 
the photon leaves the splitter (certainly, as far as the photon is 
not detected, it is always possible for the physicist to let the 
photon pass to a further interferometer and to oblige it to change 
the outcome distribution). We {\em discard} also the hypothesis that the 
values produced by photon 1 depend on the time at which photon 2 impacts
 at a detector $D_{2}$, because there is experimental evidence against 
 it \cite{com11}.

\section{Consequences}

We begin by introducing some notation: an experiment $e$ will be labeled 
by indicating the kind of impact 
that each particle undergoes, f.i. $e=(a_{1},b_{2})$. Expressions like 
$P(e_{\sigma\omega})$, $\sigma,\omega\in\{+,-\}$, denote the probabilities 
to obtain the indicated detection values in experiment $e$ (i.e., photon 1 is detected 
in $D_{1}(\sigma)$, photon 2 in $D_{2}(\omega)$). In a similar way, we 
write $P^{QM}(e_{\sigma\omega})$ for the probabilities predicted by 
standard QM for experiment $e$ (note that in this case the impacts 
can only be $u_{i}$ or $d_{i}$, since QM doesn't consider differences 
in timing).

{\em Principle II} implies that
\ba
P((a_{1},b_{2})_{\sigma\omega})=P((b_{1},a_{2})_{\sigma\omega})=\nonumber\\
=P^{QM}((u_{1},u_{2})_{\sigma\omega}).
\label{eq:p2ex}
\ea
In all interference experiments performed till now both splitters were at rest, 
and one of the impacts did happen always before the other. Then, according to 
Equation (\ref{eq:p2ex}), AD reduces to QM for all experiments already done.

On the other side, it follows from {\em Principle I} that
\ba
P((b_{1},b_{2})_{\sigma\omega})=P^{QM}((d_{1},d_{2})_{\sigma\omega})=\nonumber\\
=P^{QM}((u_{1},d_{2})_{\sigma\omega})=P^{QM}((d_{1},u_{2})_{\sigma\omega}).
\label{eq:p1ex}
\ea
Experiments in which both impacts are {\em non-before} events, or in which 
the photons impact succesively in several beam-splitters, as well as the 
generalization to n-particle experiments are discussed in other articles.

\begin{figure}[ht]
\PSpetitGr{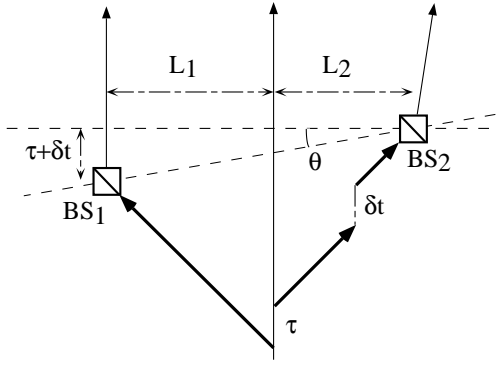}{0.4}
{\small\it{\caption{Experiment with one moving BS: time diagram in laboratory 
frame.}}}
\label{fig:timediag}
\end{figure}

Suppose now that the state of movement of the beam-splitters implies 
the following situation: The impact at $BS_{1}$, in the inertial frame of 
$BS_{1}$, occurs before the impact at $BS_{2}$, and the impact at $BS_{2}$, in the 
inertial frame of $BS_{2}$, occurs before the impact at $BS_{1}$. The diagram 
in Fig.2 
corresponds to such a {\em gedanken Experiment}: It is assumed 
that the photons are channeled from the source to the beam-splitters 
by means of optical fibers, and that the optical path $S$-$BS_{2}$ traveled 
by photon 2, is a bit longer than optical path $S$-$BS_{1}$ traveled by 
photon 1. The delay in time resulting from this path difference 
is labeled $\delta t$. At the moment photon 1 arrives at $BS_{1}$, this splitter 
is at rest, at a distance $L_{1}$ from the source. At the moment photon 
2 arrives at $BS_{2}$, this splitter is at distance $L_{2}$ from the source 
and moving with velocity $V$ in the indicated direction. The delay 
between the emissions of the two photons is labeled $\tau$.

As said, the quantum formalism does not depend at all on the inertial 
frames of the beam-splitters. The correlation coefficient is assumed 
to be given by the Lorentz-invariant expression \cite{bel64,div}:
\be
E=\sum_{\sigma,\omega}\sigma\omega 
P^{QM}((u_{1},u_{2})_{\sigma\omega})=\cos 2(\alpha+\beta).
\ee
Consequently, for $\alpha+\beta=0$, QM predicts 
perfectly correlated results (either both photons are transmitted, or 
they are both reflected).

{\em Principle I} implies that each photon produces values according 
to local information only, and equation (\ref{eq:p1ex}) leads to a 
correlation coefficient
\be
E=\cos 2\alpha\cos 2\beta.
\ee
Consequently, according to AD, $\alpha+\beta=0$ does not imply $E=1$. In 
particular, if $\alpha=-\beta=45^{\circ}$ one gets $E=0$, i.e. the four 
possible outcomes $(+1,+1)$, $(+1,-1)$, $(-1,+1)$, $(-1,-1)$ equally 
distributed.

In summary, for the {\em gedanken Experiment} of Fig.2, 
QM (according to which indistinguishability is a sufficient 
condition for superposition) and AD (according to 
which superposition does also depend on the timing of the impacts at 
the splitters) lead to clearly conflicting predictions.

\section{Possibility of a real experiment}

What about the possibility of doing a real experiment allowing us to 
distinguish between the two descriptions? The condition, guaranteeing 
that the impact of photon 2 at $BS_{2}$ occurs in the inertial frame 
of $BS_{2}$ {\em before} the impact of photon 1 at $BS_{1}$, can be 
easily derived from the diagram of Fig.2:
\be
\tan\theta=\frac{V}{c}=\frac{c(\tau+\delta t)_{max}}{L}
\label{eq:cond}
\ee
where $V$ is the velocity of $BS_{2}$, $c$ is the speed of light, 
$L=L_{1}+L_{2}$ and $(\tau+\delta t)_{max}$ is the maximal delay 
between the two impacts. For photon produced by down-conversion, 
$\tau<<\delta t$, so equation (\ref{eq:cond}) imposes the following 
condition to the value of $\delta t$:
\be
\delta t<\frac{VL}{c^2}.
\ee
Velocities of about 100 m/sec (360 km/h) can be reached by setting $BS_{2}$ on 
a fast moving wheeler (note that, according to AD, detectors don't 
need to move). 
Bell-experiments with over 4 km of optical fiber have been done 
\cite{rar94}: 
an experiment with such a value for the distance $L$ between the 
beam-splitters would allow us an upper limit for parameter $\delta t$ of 4.4 ps. 
A quantum channel of 24 km, has already been achieved \cite{gis95}: 
an experiment with such a value of $L$ would allow us an upper limit 
for $\delta t$ of 26.4 ps. The ongoing effort to achieve quantum cryptography 
over long distances, may make possible values of $L$ greater than 100 km 
in near future, which would mean that $\delta t$ could reach values of 111 ps. 
The feasibility of the relativistic experiment depends, therefore, 
basically on the capability to ensure a significant number of impacts 
respecting such limits for $\delta t$. Work to clarify whether the control of 
the setup's geometry can be implemented to this extent with reasonable 
effort is in progress.

\section{Conclusion}
In conclusion, the nonlocality of quantum mechanics is so important and 
counterintuitive that as many experiments as possible should be performed 
to get deeper insight in its nature. If the proposed relativistic 
experiment with moving beam-splitters can be carried out and the 
results uphold the predictions of the conventional superposition 
principle, then it will be hard to maintain the belief of the 
"pacific coexistence" between relativity and quantum mechanics. Then 
models giving up relativity of space-time like Bohm's causal model 
\cite{boh}, 
should be explored more in depth. On the contrary, if the results of 
the experiment with moving beam-splitters prove to be in conflict 
with quantum mechanics, the road to a new description of physical 
reality would be opened. This description would base on two main 
assumptions: (1) there are in nature faster than-light-influences 
which cannot be used by man for superluminal signaling, and (2) 
superposition depends not only on indistinguishability but also 
on the timing of the impacts at the beam-splitters. Such a description 
would be perfectly coherent with both, quantum nonlocality and space-time 
relativity. The realization of experiments allowing us to investigate 
time orderings different of the conventional ones appears promising 
in any case \cite{com14}.

\section*{Acknowledements}
We acknowledge Anton Zeilinger (University of Innsbruck) for 
discussions about possible experimental realizations. A. Suarez 
acknowledges support of the Leman and Odier Foundations.


\begin{thebibliography}{14}
\bibitem{ghz93} D.M. Greenberger, M.A. Horne, A. Zeilinger, {\em 
Physics Today}, August (1993) 24.
\bibitem{bel64} J.S. Bell, {\em Physics}, {\bf 1} (1964) 195-200; 
{\em Speakable and unspeakable in quantum mechanics}, Cambridge: 
University Press, 1987.
\bibitem{ghz89} D.M. Greenberger, M.A. Horne, A. Zeilinger, in: M. 
Kafatos (ed.), {\em Bell's Theorem, Quantum Theory, and Conceptions 
of the Universe}, Dordrecht: Kluwer Academic, 1989, p.173.
\bibitem{har93} L. Hardy, {\em Phys.Rev.Lett.} {\bf 71} (1993) 
1665.
\bibitem{div} A. Aspect, P. Grangier and G. Roger, {\em 
Phys.Rev.Lett.}, {\bf 49} (1982) 91-94; A. Aspect, J. Dallibard and G. Roger, {\em 
Phys.Rev.Lett.}, {\bf 49} (1982) 1804-1807; J.G. Rarity and P.R. 
Tapster, {\em Phys.Rev.Lett.}, {\bf 64} (1990) 2495-2498; E. Santos, {\em 
Phys.Rev.Lett.}, {\bf 66} (1991) 1388-1390 and {\em Phys.Rev.Lett.}, {\bf 
68} (1992) 2702-2703.
\bibitem{kwi94} P.G. Kwiat, P.H. Eberhard, A.M. Steinberg and R.Y. 
Chiao {\em Phys.Rev.A}, {\bf 49} (1994) 3209-3220.
\bibitem{fra89} J.D. Franson, {\em Phys.Rev.Lett.}, {\bf 62} 
(1989) 2205-2208.
\bibitem{per95} A. Peres, in: D.M. Greenberger, M.A. Horne, A. 
Zeilinger (eds.), {\em Fundamental Problems in Quantum Theory}, New 
York: New York Academy of Sciences, 1995, p.445.
\bibitem{boh} D. Bohm, {\em Phys.Rev.}, {\bf 85} (1952) 166-193; 
D. Bohm and B.J. Hilley, {\em The Undivided Universe}, New York: 
Routledge, 1993.
\bibitem{com10} To our knowledge the distance splitter-detector has not 
been considered a relevant parameter in any performed experiment. 
Nevertheless the independence of the outcomes on this distance can 
be considered demonstrated by the fact that correlations have been 
observed as well by experiments using distances splitter-detector 
of some cm (see for instance A.R. Wilson, J. Lowe and D.K. Butt, 
{\em J. Phys. G: Nucl. Phys.}, {\bf 9} (1976) 613-624), as by others using 
distances of 1 m (for instance the experiment by J.G. Rarity and P. R. Tapster
referred to in \cite{div}, private communication 22.9.1995).
\bibitem{com11} In the experiment by Rarity and Tapster referred to in 
\cite{div}, the impact of each photon at the {\em beam-splitter} occurs 
{\em before} the impact of the other photon at the {\em detector}. 
Were in {\em Principle I} the term "before" defined with relation 
to the impact at a detector $D_{2}$ instead to the impact at $BS_{2}$, 
the model would predict the dissapearance of the nonlocal correlations, 
and contradict the observed results.
\bibitem{rar94} J.G. Rarity and P.R. Tapster, {\em Phys.Rev.Lett.}, {\bf 
73} (1994) 1923-1926.
\bibitem{gis95} A. M\"{u}ller, H. Zbinden and N. Gisin, {\em Nature}, {\bf 
378} (1995) 449.
\bibitem{com14} The study of other time sequences using experiments in which 
one of the photons impacts successively at two beam-splitters, as well as a 
detailled and generalized presentation of the proposed alternative description 
is carried out by one of us in: A. Suarez, {\em Nonlocal phenomena: A 
superrelativistic description with many entanglement rules}, preprint CQP-960107; 
and A. Driessen and A. Suarez (eds.),{\em Nonlocal phenomena: physical explanation 
and philosophical implications}, Dordrecht: Kluwer, 1997, in press.
\end{thebibliography}
\end{document}